\documentclass[twocolumn,twocolappendix]{aastex63}

\usepackage{amsmath}
\usepackage{nccmath} 
\usepackage{amssymb}
\usepackage{bm}
\usepackage{graphicx}
\usepackage{color}
\usepackage{float}
\usepackage{multirow}
\usepackage{CJK}
\usepackage[T1]{fontenc}
\usepackage[hang,flushmargin]{footmisc}
\usepackage{pifont}

\usepackage{xcolor}
\usepackage{ulem} 

\definecolor{addcolor}{rgb}{0.8, 0.0, 0.0}   
\definecolor{delcolor}{rgb}{0.0, 0.0, 0.8}   
\definecolor{modcolor}{rgb}{0.0, 0.0, 0.8}   
\definecolor{commentcolor}{rgb}{0.6, 0.4, 0.0} 

\begin{document}

\begin{CJK*}{UTF8}{ipxm}
\title{Planet Migration in Protoplanetary Disks with Rims}

\author[0009-0004-7622-3056]{Zhuoya Cao（曹卓雅）}
\affiliation{Department of Astronomy, Westlake University,
Hangzhou, 310024, China}
\affiliation{Department of Astronomy, Tsinghua University, Beijing, 100084, China}

\author[0000-0002-7329-9344]{Ya-Ping Li \begin{CJK*}{UTF8}{gbsn}(李亚平)\end{CJK*}}
\affiliation{Shanghai Astronomical Observatory, Chinese Academy of Sciences, Shanghai 200030, China}

\author[0000-0001-5466-4628]{Douglas N.C. Lin（林潮）}
\affiliation{Department of Astronomy and Astrophysics, University of California, Santa Cruz, CA 95064, USA}
\affiliation{Institute for Advanced Studies, Tsinghua University, Beijing, 100084, China}
\affiliation{Department of Astronomy, Westlake University, Hangzhou, 310024, China}

\author[0000-0001-8317-2788]{Shude Mao（毛淑德）}
\affiliation{Department of Astronomy, Westlake University, Hangzhou, 310024, China}

\correspondingauthor{Yaping Li}
\email{liyp@shao.ac.cn}


\begin{abstract}
Complex structures, including sharp edges, rings and gaps, have been commonly observed in 
protoplanetary disks with or without planetary candidates. Here we consider the possibility that they
are the intrinsic consequences of angular momentum transfer 
mechanisms, and investigate how they may influence the dynamical
evolution of embedded planets.  
With the aid of numerical hydrodynamic simulations, we show that gas giants have a tendency to migrate away from sharp edges, whereas super-Earths embedded in the annuli tend to be retained. This implies that, observationally, Jupiters are preferentially detected in dark rings (gaps), whereas super-Earths tend to be found in bright rings (density bumps). 
Moreover, planets' tidal torque provide, not necessarily predominant, feedback on the surface density profile.  
This tendency implies that Jupiter's gap-opening process deepens and widens the density gap associated with the dark ring, while super-Earths can be halted by steep surface density gradient near the disk or ring boundaries. 
Hence, we expect there would be a desert for super-Earths in the surface density gap.




\end{abstract}


\section{Introduction}







Protoplanetary disks (PPDs) are residuals of star formation and provide the initial environment for planetary assembly, setting the stage for the demographics of mature planetary systems \citep{Hartmann2001}. Planetary formation and evolution are intrinsically linked to tidal interactions with the protoplanetary disk \citep{ida2004, Armitage2011}. Low-mass planets, including super-Earths and Neptune-mass planets, usually exert only a modest perturbation on their disk's structure. The net torque on low mass planets from the disk drives rapid type I migration
\citep{goldreich1980,tanaka2002,paardekooper2011, Lubow2010, tanaka2002, McNally2019, Ward1997}. Massive planets strongly perturb their disk and open deep gaps in the gaseous disk through gravitational torques exerted at Lindblad and co-rotation resonances 
\citep{lin1986a, Calvet2002}, facilitating angular momentum transport in a 
viscous-like manner \citep{lin1986b, lin1993}. This is the type II regime \citep{Nelson2000, Lubow2006, Drmann2015, Dong2015a, Kanagawa2018, Robert2018}. Both processes can drive the efficient inward migration of the embedded planet, which could explain
the origin of close-in planets \citep{lin1996}.

Recent hydrodynamical simulations indicate that migration rates in the type I and type II regimes can be described by a more generalized migration prescription after considering the gap-opening effect \citep{Kanagawa2018}. 
The migration rate from these simulations, even after considering the suppression due to the gap-opening effect, still poses a challenge to the retention of cold Jupiters when compared with 
the observed period distribution \citep{Nelson2000,Ida2013}. 
Many scenarios have been explored to identify processes which may lead to 
the slowing down of migration speed or even the reversal of migration direction \citep[e.g.,][]{Masset2006,Lega2014,Chen2020,Dempsey2021, li2024,Wu2024,Laune2024,Hou2024,Wu2025,Liu2025,Pan2026,Ida2026}.



High-resolution ALMA observations of the protoplanetary disk have revealed a series of concentric ring- and gap-like substructures (bright rings and narrow dark annuli/gaps) in the dust continuum emission \citep[e.g.,][]{Andrews2016,Andrews2018,Long2018,Segura2020,Guidi2022,Booth2024,Carvalho2024,Teague2025,Sierra2025,Gardner2025}. 
Such kind of sub-structures in the dust continuum could be attributed to the intrinsic consequences of angular momentum transfer mechanisms like the dead zone \citep{Bai2013, Lyra2008, Li2019}, magnetohydrodynamics zonal flows \citep{Bai2014, Johansen2009, Krapp2018, Cui2021}, magnetized disk winds \citep{Suriano2017, Riols2019}, vertical shear instability \citep{Nelson2013, Lin2015}, the eccentric cooling instability \citep{Lubow1991, LiJ2021}, secular gravitational instability \citep{Youdin2011, Tominaga2019}, or induced by the embedded planets \citep{lin1986b,Dong2015a,Zhang2018,Li2019b,Paardekooper2023}. 
Some kinematic evidence for protoplanets forming in ringed-structure disks has emerged from the detection of localized deviations from Keplerian velocity in HD 163296 \citep{Pinte2018, Teague2018}.

Most studies on planet-disk interaction have focused on disks with smooth initial surface density profiles. As the annular structures such as rings and gaps could coexist in PPDs, the migration pattern in these disks could be important to shape the architectures of exoplanet system, which are, however, remains relatively unexplored. \citet{Masset2006} found that a reversed gas surface density gradient, for instance near a cavity, can generate a large, positive corotation torque for low-mass planets, effectively halting their inward migration. This study establishes a new framework for investigating planetary reverse migration and migration arrest \citep[see also,][]{Liu2025}. 


In this work, we investigate the planet migration with rims in disks, including gas bumps and gaps, for various planetary masses using hydrodynamical simulations. We investigate the mechanism by setting an $\alpha$ dead zone or viscosity bump profile, which can lead to the disk surface density redistribution. 

The rest of the paper is organized as follows. In the Methods section (\S \ref{sec:methods}), we describe the numerical setup for the hydrodynamical simulations. We show the simulation results and implications in \S \ref{sec:results}. The conclusion of this work is presented in \S \ref{sec:discussion}.


\section{Methods}
\label{sec:methods}
In this section, we describe the simulation setup and diagnostics for the migration of the planet with a disk rim. 
To investigate the influence of the disk on planetary migration, we adopt the conventional $\alpha$-prescription
for steady-state accretion disk models \citep{ShakuraSunyaev1973} and
modify the $\alpha$ profile to alter the disk's surface density distribution. A surface density bump can be achieved by implementing a dead zone in the $\alpha$ profile, while a surface density gap can be produced by establishing an $\alpha$ bump configuration. 



\subsection{Simulation Setup}

We globally simulate the hydrodynamical evolution of a planet embedded in a PPD with \texttt{Athena++} \citep{Stone2020}.
The continuity equations for the mass and momentum of the fluid are numerically solved in a 2D thin disk by including the gravitational interaction between the planet and the disk. 
The coordinate of the system is located at the position of the central star with its mass of $M_{*}$. For the 2D disk model, we use the cylindrical coordinate $(r,\ \phi)$ for the simulations.  

We follow the same method as in previous works to initialize the disk \citep[e.g.,][]{Li2019,Liu2025}. The disk surrounding a pre-main sequence (PMS) star of mass $M_\star = 1.0 M_\odot$ is initialized with a gas surface density profile following a power-law distribution, which is described as

\begin{equation}
\Sigma(r)=\Sigma_0(r_0)\left(\frac{r}{r_0}\right)^{-0.5}, 
\end{equation}
where $r_{0}$ is the typical scale length of the disk. The locally isothermal temperature profile is
initialized as
\begin{equation}
T(r)=T(r_0)\left(\frac{r}{r_0}\right)^{-1}.
\end{equation}
This disk temperature profile leads to a constant disk aspect ratio with $h\equiv H/R$, where $H$ denotes the disk scale height. Here we adopt a constant disk aspect ratio of $h=0.05$.






\subsubsection{The Disk Rim Model}






We adopt a transition of the disk viscosity (i.e., the dead zone) to approximate the disk rim model. 
The dead zone in PPDs is closely related to the magneto-rotational instability (MRI) \citep{Balbus1991}, which was long thought to govern accretion in protoplanetary disks. However, low ionization rates in these disks are expected to suppress the MRI within a region of $\sim$1–10 AU, leading to significantly reduced turbulence and angular momentum transport compared to theoretical expectations \citep{Lehmann2022, Gressel2015}, which is known as the ``dead zone'' \citep{Gammie1996, Turner2009, Armitage2011, Turner2014}.
The formation of this dead zone is attributed to a low ionization fraction, which prevents the gas from coupling effectively with the magnetic field, a dependency that is further influenced by the dust properties within the disk \citep{Dzyurkevich2013}. The specific location and radial extent of the dead zone are highly uncertain, as they are governed by complex factors including the sources of ionization, the disk's physical structure, and the local magnetic field strength.

The dead zone typically manifests as a sharp viscosity gradient at a distance of a few astronomical units from the central star \citep{Armitage2011}. In fact, recent state-of-the-art magnetohydrodynamical models show that the inner parts of protoplanetary disks (between $\sim$ 1-20 AU) are largely laminar, with angular momentum transport driven by magneto-thermal winds and laminar Maxwell stresses \citep{Gressel2015, Bai2015}. 

\begin{figure*}
    \centering
    \includegraphics[width=0.95\linewidth]{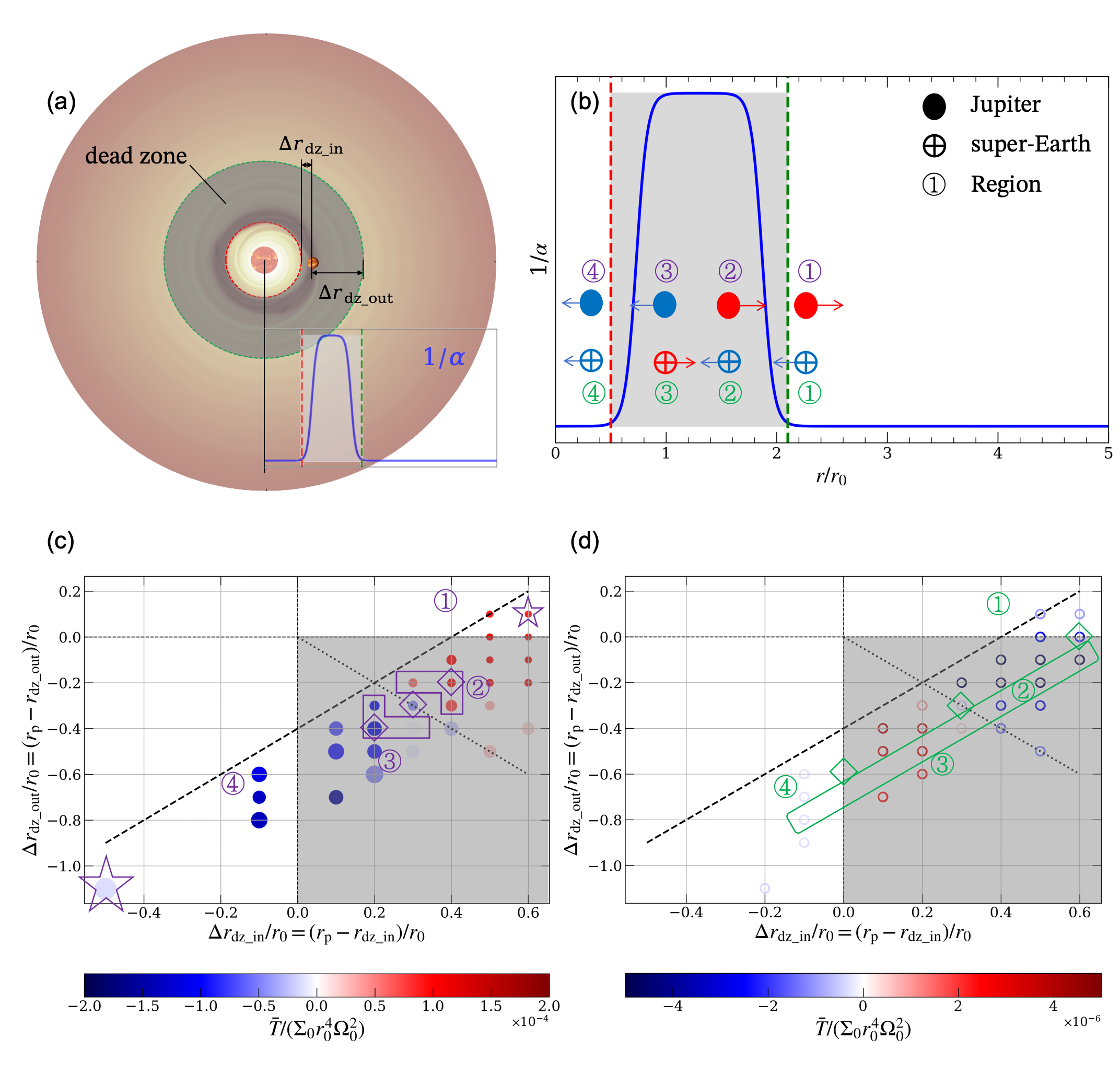}
    \caption{Schematic of dead zone structure (panel a), planetary migration directions (panel b), and torque distribution relative to dead zone boundaries (panels c and d). Panel a is a conceptual graph of the disk and dead zone. The gray region in the figure denotes the dead zone (with initial $\alpha_{\rm dead} = 10^{-4}$ for Jupiter and $\alpha_{\rm dead} = 10^{-3}$ for super-Earth) , while the remaining disk regions have initial $\alpha_{\rm active} = 10^{-2}$. Panel b illustrates the migration directions of Jupiters and super-Earths in a protoplanetary disk with a surface density bump. The solid circle denotes Jupiter, while the symbol "$\bigoplus$" denotes the super-Earth. Blue indicates inward migration, and red indicates outward migration. In panels c and d, the x-axis represents $\Delta_{\rm dz\_in}$, defined as the distance from the planet to the inner boundary of the dead zone, and the y-axis represents $\Delta_{\rm dz\_out}$, defined as the distance from the planet to the outer boundary of the dead zone. Symbols enclosed by numbers indicate different regions where the planet is located relative to the ring/gap. 
    The color of the points corresponds to the torque value: blue points represent negative torque, corresponding to inward migration, and red points represent positive torque, corresponding to outward migration.  
    }
    \label{fig:torque_scatter}
\end{figure*}

The gas viscosity is numerically modeled as $\alpha$-prescription with $\nu=\alpha H^{2}\Omega$ \citep{ShakuraSunyaev1973}, where $\Omega$ is the angular velocity. 
To implement the dead zone in the disk, we adopt an $\alpha$ profile as \citep{Pinilla2016}

\begin{equation}
\begin{aligned}
\alpha(r) &= \alpha_{\text{dead}} + (\alpha_{\text{active}} - \alpha_{\text{dead}}) \\
&\medmath{\cdot 
\left[ 
\frac{1}{2} \left(1 - \tanh \frac{(r - r_{\text{dz\_in})}}{\delta} \right) 
+ 
\frac{1}{2} \left(1 + \tanh \frac{(r - r_{\text{dz\_out})}}{\delta} \right) 
\right]}.
\label{eq:alpha_profile}
\end{aligned}
\end{equation}
The configuration for the disk viscosity profile has a gap in the $\alpha$ profile with $\alpha=\alpha_{\rm dead}$ between $r_{\rm dz\_in}$ and $r_{\rm dz\_out}$, while it remains constant for the background disk. 
The parameter $\delta$ controls the transition slope of the viscosity between the dead zone and the active zone. This viscosity profile can mimic the coupled effect of the dead zone and the MHD wind \citep{Pinilla2016}. For our Jupiter-mass planet, $\alpha_{\rm dead} = 10^{-4}$ to amplify the effect of the dead zone, but $\alpha_{\rm dead} = 10^{-3}$ for the super-Earth mass planet. We have also tested $\alpha_{\rm dead} = 10^{-3}$ for the Jupiter-mass planet, which does not change our results significantly. For both cases, the background viscosity parameter is set to $\alpha_{\rm active} = 10^{-2}$. The conceptual graph is shown in Fig.~\ref{fig:torque_scatter}(a). 
Although the $\alpha$-profile is hypothesized in this work, this is simply used to produce the disk rim and is well motivated by numerical simulations and high-resolution observations, as mentioned above. Different models generating the same disk rim should not change our main results (see also discussions in Section~\ref{sec:torque_bump}).

\begin{figure}
    \centering
    \includegraphics[width=0.8\linewidth]{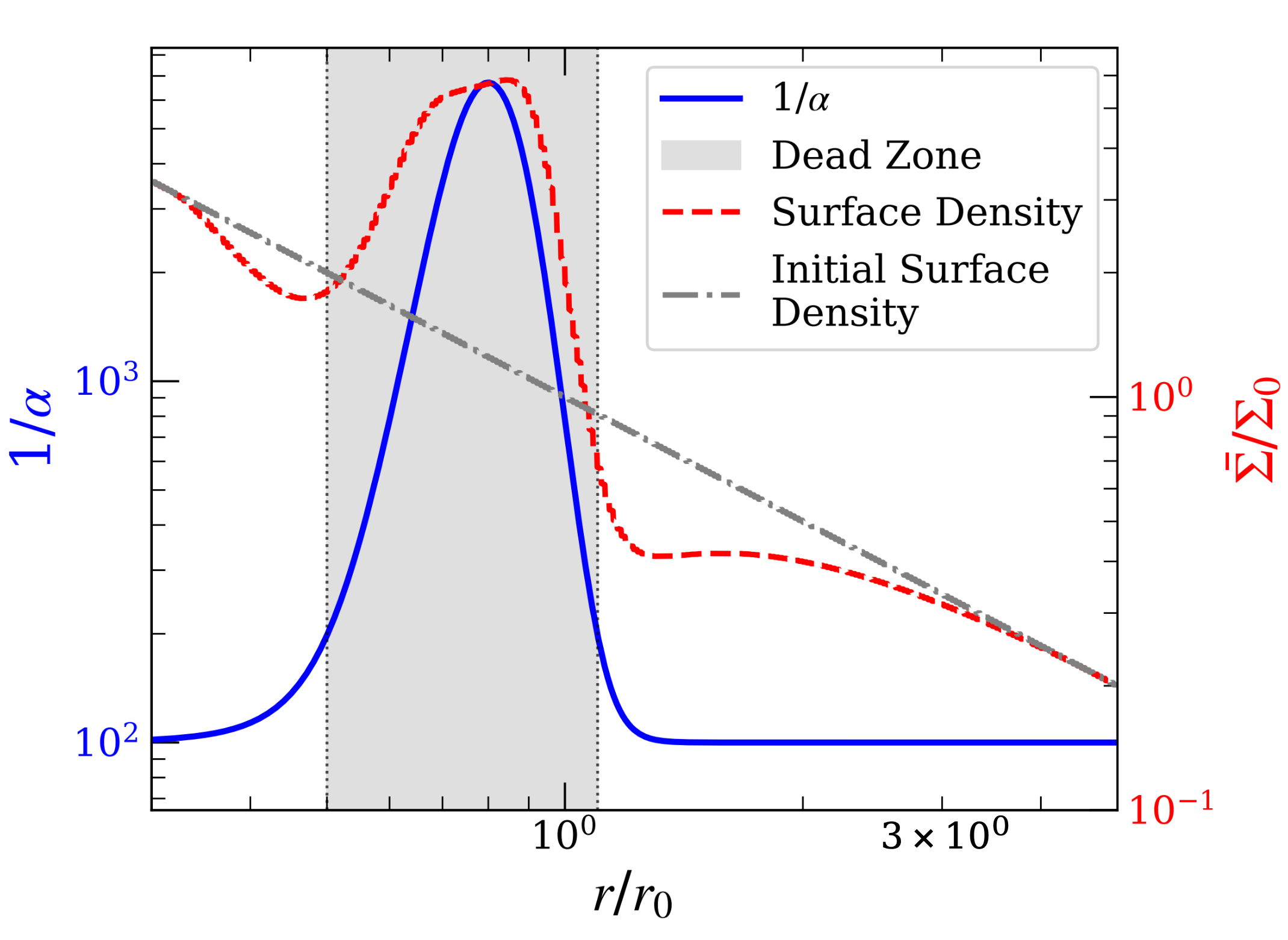}
    \caption{The azimuthally averaged 
    surface density profiles and $\alpha^{-1}$ profile of the dead zone ($\alpha_{\rm dead}=10^{-4}$) in the absence of planets. 
    The red dashed curve represents the gas surface density profile after 1000 orbits while the blue solid line corresponds to $1/\alpha$. By applying a dead zone in the $\alpha$ profile, we obtain an annular surface density bump in the resulting gas surface density. The surface density peak aligns with the region of reduced $\alpha$ in the dead zone. 
    } 
    \label{fig:no_planet}
\end{figure}

For a disk without planets, the initial state is shown in Fig.~\ref{fig:no_planet}. The $1/\alpha$ distribution follows Eq.~\ref{eq:alpha_profile} with $\delta = 0.15$. The azimuthally averaged radial profile of the gas surface density is also displayed. A surface density bump can be seen within the dead zone, corresponding to the region of reduced $\alpha$, indicating the accumulation of material due to suppressed turbulence.

\subsubsection{Planet-Disk Interaction}










The gravitational potential of the planet at position $\boldsymbol{r}$ is calculated using a smoothed potential of the form:

\begin{equation}\label{eq:pot}
\phi_{\rm p} = -\frac{GM_{\rm p}}{\left(|\boldsymbol{r}_{\rm p} - \boldsymbol{r}|^2 + \epsilon^2 \right)^{1/2}} + q\Omega_{\rm p}^2\boldsymbol{r}_{\rm p}\cdot\boldsymbol{r},
\end{equation}
where $\boldsymbol{r}_{\rm p}$ denotes the position vector of the planet, $\epsilon$ is the softening length and $q$ represents the mass ratio $M_p/M_{\odot}$. The second term on the right-hand side represents the indirect term arising from the use of a heliocentric coordinate system.
We have considered two different planet masses. One is for the Jupiter case with the planet mass set to $M_{\rm p}=0.001\ M_{\odot}$, another is for the super-Earth case with its mass of $M_{\rm p}=3\times10^{-5}\ M_{\odot}$. The softening length is set as $\epsilon=0.6\ H$ for the super-Earth case, and a larger softening length $\epsilon=0.8\ H$ is set for the Jupiter case to smooth out the fluctuated torque around the planet. Both of them are very close to the calibrated softening length from the 3D simulations \citep{Muller2012,Cordwell2025}.
The planet is fixed in a circular orbit at a distance $r_{\rm p} = r_0$ with $\Omega_{\rm p}=\Omega_{0}$, where $\Omega_{0}$ is the Keplerian frequency at $r_{0}$. 

We also explore the effect of accretion on the planet migration with a dead zone.
For the active accretion of an embedded planet, we adopt a sinkhole approach around the planet, following previous studies \citep{Li2023, li2024}. The accretion process is governed by the sinkhole radius $r_a$ and a mass removal rate $\eta$, expressed in units of the local Keplerian frequency $\Omega_0$. Within each numerical time step, a uniform fraction of mass is removed from every grid cell located within a distance $\delta r < r_a$ (where $\delta r = |\boldsymbol{r} - \boldsymbol{r}_{\rm p}|$) from the accretor. This procedure ensures that once the surface density profile inside the sinkhole reaches a steady state, the mass removal rate converges to the integrated mass flux into the sinkhole. 
In our accreting planet setup, we use the parameters $r_a = 0.1\ r_{\rm H}$, $\eta = 10\Omega_0$ where $r_{\rm H}=r_{\rm p}(M_{\rm p}/3M_{*})^{1/3}$ is the Hill radius, and the softening length is the same as the accretion radius.

The disk extends from an inner radius of $r_{\text{in}} = 0.3r_{0}$ to an outer radius of $r_{\text{out}} = 5r_{0}$. Our computational grid employs both radial ($n_{r}=240$) and azimuthal($n_{\phi}=512$) discretization, with three levels of mesh refinement implemented around the planet. This refinement achieves a resolution of $\Delta r \approx 0.001$ in both directions at $r = r_{0}$. For each configuration, we simulate 1000 to 2000 planetary orbits until the gravitational torque profile becomes stable.


\subsection{Data Analysis}

\subsubsection{Total Torque Calculation}

By varying the location of the dead zone boundary, we compute the torque exerted on the planet to determine its migration behavior at different positions relative to the surface density profile bump. 

The gravitational torque on the planet is computed by direct summation of cell-wise contributions across the computational domain. 
The softened gravitational force from each fluid element is given by:
\begin{equation}
\mathbf{F}_{ij} = \frac{GM_{\rm p} dm_{ij}}{(d_{ij}^2 + \epsilon^2)^{3/2}} \mathbf{d}_{ij}, 
\end{equation}
where $G$ is the gravitational constant, $d_{ij}$ is the distance between each grid cell and the planet, $dm_{ij}$ is the cell mass. Here we have neglected the indirect term in Eq.~\ref{eq:pot}, which is negligible. The total torque about the coordinate origin is then integrated as:
\begin{equation}\label{eq:Ttot_sim}
T = \sum_{i,j} \left( \mathbf{r}_p \times \mathbf{F}_{ij} \right)_z. 
\end{equation}
The torque averaged in time is obtained by averaging over the last 160 orbits.  

\subsubsection{Type I Migration Torque}
\label{sec:type_1_theory}

For super-Earths, semi-analytical formula for type I migration can be used to calculate the Lindblad torque and the corotation torque. 

The gravitational potential of the planet can be expanded in a Fourier series as \citep{Kley2012}:

\begin{equation}
\begin{aligned}
    \psi_{\rm p}(r,\varphi,t)&\equiv-\frac{GM_{\rm p}}{\sqrt{|\vec{r}_{\rm p}(t)-\vec{r}|^2+\epsilon^2}}\\
    &=\sum_{m=0}^\infty\psi_m(r)\cos\{m[\varphi-\varphi_{\rm p}(t)]\}, 
\label{eq:fourier}
\end{aligned}
\end{equation}
where the expression for the coefficient $\psi_m$ is given by \citep{Meyer1987}.

\begin{equation}
\begin{aligned}
    \psi_m(r) &= -\frac{G M_{\rm p}}{a_{\rm p}}\cdot b_{1/2}^{(m)}\left(\frac{r}{a_{\rm p}}, \varepsilon\right)\\
    b_{1/2}^{(m)}(\alpha, \varepsilon) &= \frac{1}{\pi} \int_0^{\pi} \frac{\cos(m\theta)}{\sqrt{1 + \alpha^2 - 2\alpha\cos\theta + \varepsilon^2}}  d\theta.
\label{eq:laplace}
\end{aligned}
\end{equation}

The torque acting on a super-Earth consists of the Lindblad torque and the corotation torque. 
The Lindblad resonance radius in a nearly Keplerian disk is given by Eq.~\ref{eq:lindblad_resonance},

\begin{equation}
    r_\mathrm{L}=\left(\frac{m}{m\pm1}\right)^{-2/3}r_{\rm p}.
\label{eq:lindblad_resonance}
\end{equation}

The Lindblad torque in each $m$ mode is given by  \citep{Chen2020},

\begin{equation}
\begin{aligned}
    \Gamma_{L,m} &= \frac{\pi^2 m \Sigma_L}{r_L\cdot dD_m/dr \Big|_{r=r_L}} \\
    ~~&\cdot \left( r \cdot \frac{d\psi_m}{dr} + 2m^2(1-\frac{\Omega_{\rm p}}{\Omega}) \psi_m\right)^2\Big|_{r=r_L} \cdot f_L,
\label{eq:lindblad}
\end{aligned}
\end{equation}
where $D_m(r) = \kappa^2-m^2(\Omega-\Omega_{\rm p})^2$, the cutoff coefficient $f_L = [\sqrt{1+m^2h^2}(1+4m^2h^2)]^{-1}$. For a Keplerian disk, $\kappa(r) = \Omega(r)$. 

The expression for the corotation torque is given by \citep{goldreich1980}, 

\begin{equation}
\begin{aligned}
    \Gamma_{C,m} = \frac{m\pi^2}{2} \cdot \frac{\psi_m^2(r_C)}{d\Omega/dr\Big|_{r=r_C}} \cdot \frac{d}{dr}\left(\frac{\Sigma}{B}\right)\Big|_{r=r_C},   
\label{eq:corotation}
\end{aligned}
\end{equation}
where $B=-1/2(\bm \nabla \times \vec{v})_z=-1/2(\partial v_{\phi}/\partial r + v_{\phi}/r-1/r\cdot\partial v_r/\partial \phi)$. 

The total torque is given by the sum of the Lindblad torque and the corotation torque, as expressed in Eq.~\ref{eq:tot}. 

\begin{equation}
\Gamma_{\rm tot}=\sum_{m=1}^{\infty}\Gamma_{L,m}+\sum_{m=1}^{\infty}\Gamma_{C,m}. 
\label{eq:tot}
\end{equation}

\begin{figure*}
    \centering
    \includegraphics[width=0.9\linewidth]{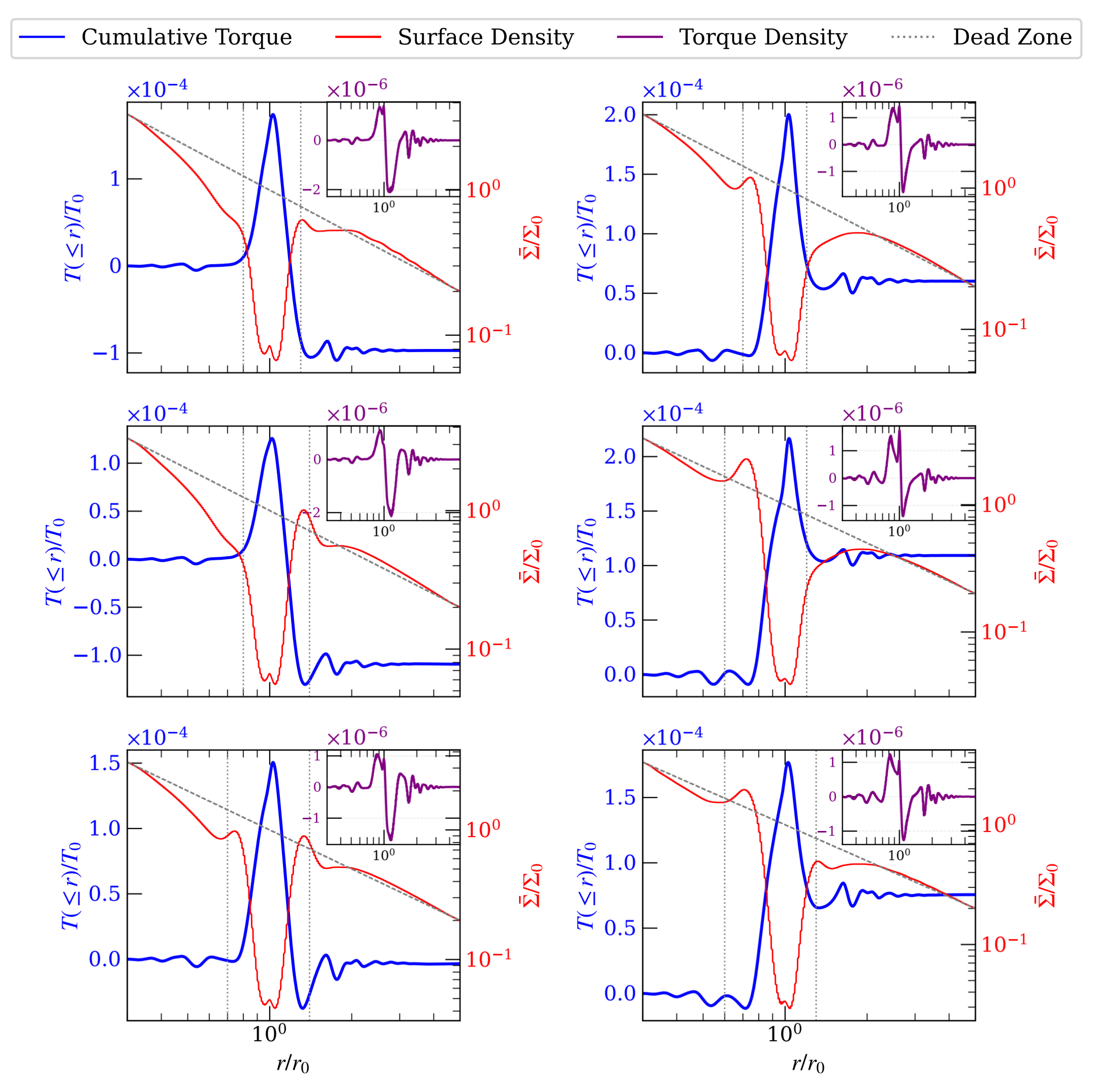}
    \caption{Radial surface density (averaged) profiles and radial torque (sum) profiles for the models with a Jupiter mass planet. Different panels correspond to different runs (enclosed by the purple `L' shaped structures) as shown in the panel c of Fig.~\ref{fig:torque_scatter}. 
    The simulation parameters are set with $M_{\mathrm{p}} = M_{\mathrm{J}}$. The inner and outer dead-zone boundaries $r_{\mathrm{dz\_in}}$ and $r_{\mathrm{dz\_out}}$ are indicated in each panel. For the left column (top to bottom): the top panel employs $r_{\mathrm{dz\_in}} = 0.8\,r_0$, $r_{\mathrm{dz\_out}} = 1.3\,r_0$; the middle panel uses $r_{\mathrm{dz\_in}} = 0.8\,r_0$, $r_{\mathrm{dz\_out}} = 1.4\,r_0$; and the bottom panel is specified as $r_{\mathrm{dz\_in}} = 0.7\,r_0$, $r_{\mathrm{dz\_out}} = 1.4\,r_0$. For the right column (top to bottom): the top panel adopts $r_{\mathrm{dz\_in}} = 0.7\,r_0$, $r_{\mathrm{dz\_out}} = 1.2\,r_0$; the middle panel has $r_{\mathrm{dz\_in}} = 0.6\,r_0$, $r_{\mathrm{dz\_out}} = 1.2\,r_0$; and the bottom panel corresponds to $r_{\mathrm{dz\_in}} = 0.6\,r_0$, $r_{\mathrm{dz\_out}} = 1.3\,r_0$.
    The red solid curves trace the simulated surface density distribution $\Sigma(r)$, while 
    gray dashed lines indicate the initial surface density slope $\Sigma_0(r)$ for reference. The dead zone is marked by gray dotted lines. The blue solid curve represents the radial cumulative torque, which is azimuthally integrated. The corresponding differential torque density profiles are shown by the purple solid lines in the inset at the upper right corner.  
    }
    \label{fig:density_r_torque_diff}
\end{figure*}

\section{Results}
\label{sec:results}



In this section, we explore migration dynamics for Jupiter and super-Earth mass planets with different disk rim structures. 
The main simulation results with a surface density bump in the disk are summarized in Fig.~\ref{fig:torque_scatter}.

\subsection{Jupiter Migration}

We first consider the case of a Jupiter mass planet, which is shown in the lower left panel of Fig.~\ref{fig:torque_scatter}.
We explore the migration dynamics of the embedded planet in the disk with rim at different locations. The location of the rim can be characterized by $\Delta r_{\rm dz\_in} = r_p - r_{\rm dz\_in}$  which indicates planet orbital distance from the inner dead zone boundary, and $\Delta r_{\rm dz\_out} = r_p - r_{\rm dz\_out}$ which measures the distance from the outer edge.
Fig.~\ref{fig:torque_scatter}(c) summarizes torque measurements from 30 Jupiter simulation configurations, with each case located in different regions characterized by the coordinate ($\Delta r_{\rm dz\_in},\Delta r_{\rm dz\_out}$).
The symbol color encodes the torque sign and magnitude: redder color suggests stronger positive torque/outward migration, while bluer color means stronger negative torque/inward migration.

There is a general trend that the planet would migrate outward when the rim is located at $\Delta r_{\rm dz\_out}>-\Delta r_{\rm dz\_in}$, while the planet would migrate inward if the rim location satisfies $\Delta r_{\rm dz\_out}
<-\Delta r_{\rm dz\_in}$ (see Panel b in Fig. \ref{fig:torque_scatter}). There is nearly 
zero migration torque when the planet is located at the rim center where $\Delta r_{\rm dz\_out}\simeq -\Delta r_{\rm dz\_in}$, shown as the dotted line in the panel c of Fig.~\ref{fig:torque_scatter}. This simply suggests the dead zone center ($r_{\rm mid} = (r_{\rm dz\_in} + r_{\rm dz\_out})/2$) as the critical boundary of the migration torque reversal.
Spatial gradients in the color map reveal smooth variations in the migration torque in terms of both its amplitude and sign. The size of the symbol reveals the variance of the torque, which could be related to the vortex-induced torque variations, as we show below.

We analyze six representative runs with the location symmetric about the $\Delta r_{\rm dz\_out} = -\Delta r_{\rm dz\_in}$ boundary: 
three in inward-migration domains (left of boundary) and three in outward-migration zones (right of boundary) (purple `L' shapes in Fig.~\ref{fig:torque_scatter}(c)). The total torque, when only linear terms are considered, is the sum of the Lindblad torque and the corotation torque. Due to gap opening by Jupiter, the disk near the planet is sufficiently depleted, rendering the corotation torque negligible. Fig.~\ref{fig:density_r_torque_diff} reveals the azimuthally averaged surface density as a function of radial distance $r$. The three panels in the left column exhibit a surface density peak outside the planet's orbit, indicating a Lindblad torque directed inward, which promotes inward migration. The three panels in the right column show a surface density peak inside the planet's orbit, implying an outward Lindblad torque that leads to outward migration. 
The cumulative torque profile (Fig.~\ref{fig:density_r_torque_diff}) reveals a systematic transition: as planetary orbital position shifts from inward-migration-dominated zones (characterized by prevalent blue features) to outward-migration-favored regions (increasingly dominated by red symbols), positive torque contributions progressively strengthen and expand their spatial influence. 


On a structured disk with a surface density bump, Jupiters tend to migrate away from the bump. Consequently, it is less likely to find a Jupiter embedded in a bright ring, which observationally traces a surface density bump.

\subsection{Super-Earth Migration}

The simulation results for the super-Earth migration are summarized in Fig.~\ref{fig:torque_scatter}(d). We will use the same boundaries for the dead zone to divide the parameter space into four regions: \ding{192}, \ding{193}, \ding{194}, and \ding{195}, as indicated by the green labels in Fig.~\ref{fig:torque_scatter}(b). Planets in regions \ding{192} and \ding{195} experience minimal influence from the surface density bump and undergo inward migration. Planets in region \ding{193} migrate inward, while those in region \ding{194} migrate outward. This convergent motion toward the bump center in the disk may ultimately cause super-Earths to be trapped within the surface density bump.  

\begin{figure}
    \centering
    \includegraphics[width=0.8\linewidth]{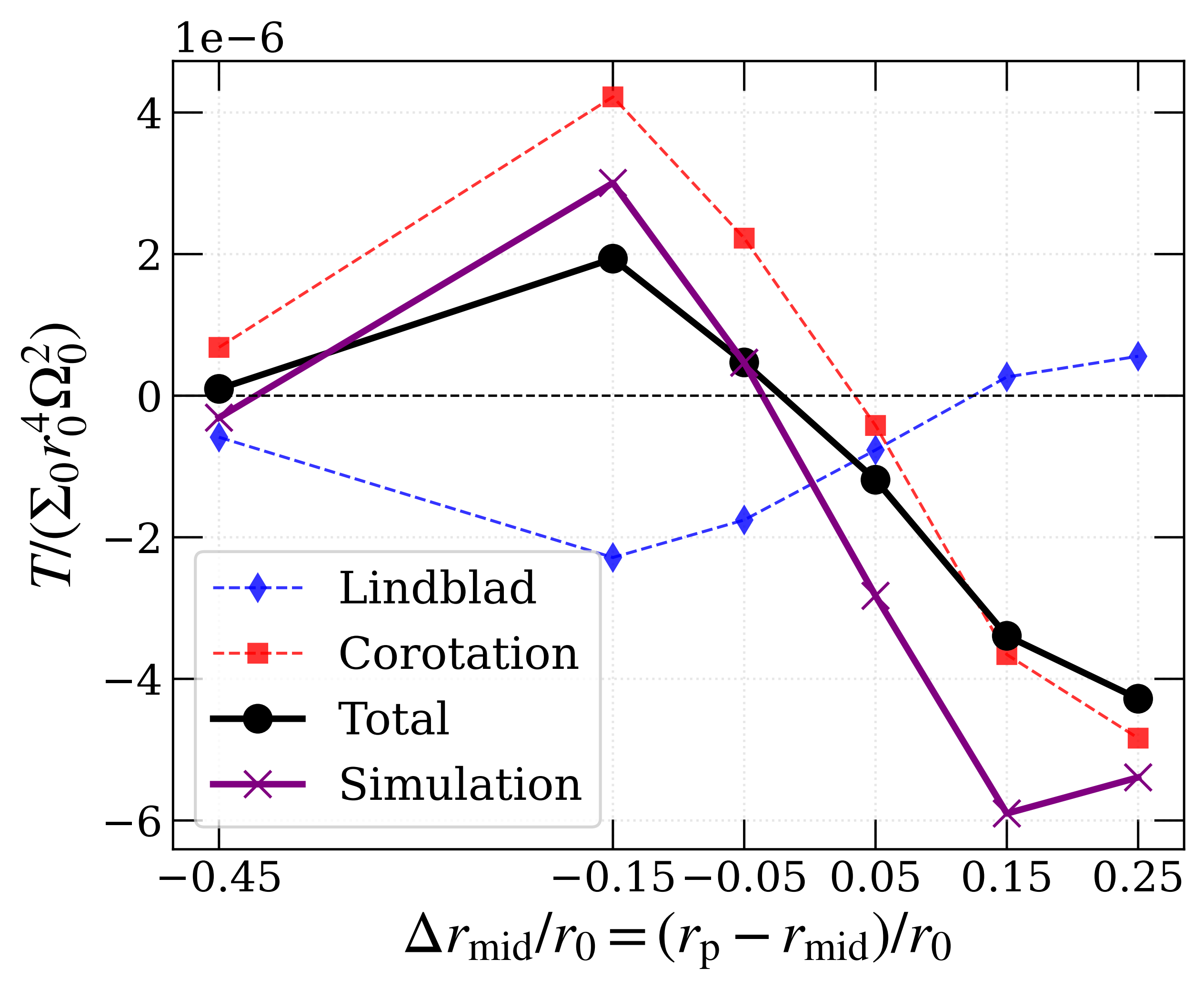}
    \caption{Theoretical calculation of the torque for super-Earth. Here, we set $M_{\rm p} = 10M_{\oplus}$. We compute the theoretical torque values for the models enclosed by the green rectangle in Fig.~\ref{fig:torque_scatter}(d). All these models have a dead zone width of $0.7r_0$, and the 
    middle line of the 
    surface density bump $r_{\rm dz\_mid}$ is shown on the horizontal axis of the figure. The blue diamond-dashed line represents the Lindblad torque, and the red square-dashed line denotes the corotation torque.  The black dot-solid line shows the sum of the two obtained from analytical calculation, representing the total torque. The purple x-solid line corresponds to the total torque directly computed from Eq.~\ref{eq:Ttot_sim}. 
    }
    \label{fig:theory}
\end{figure}

For super-Earths, we can apply semi-analytical calculations by the formula for type I migration (Section \ref{sec:type_1_theory}). The results for the comparison between theoretical and simulated torque values of the models enclosed by the green rectangle in Fig.~\ref{fig:torque_scatter}(d) are illustrated in Fig.~\ref{fig:theory}. 

As shown in Fig.~\ref{fig:theory}, the blue diamond-dashed line represents the Lindblad torque, the red square-dashed line denotes the corotation torque, and the black dot-solid line shows the total torque. The purple solid line with the cross symbols corresponds to the total torque directly computed using Eq.~\ref{eq:Ttot_sim}. Total torques calculated from theory and simulation are generally consistent with each other in both variation trend and order of magnitude. However, certain discrepancies may exist near the sign transition boundaries. The variation of the corotation torque (red curve) closely follows that of the total torque, indicating that for super-Earths, the corotation torque plays a dominant role in determining the migration direction, while the Lindblad torque has a relatively minor influence. 

\begin{figure*}
    \centering
    \includegraphics[width=0.9\linewidth]{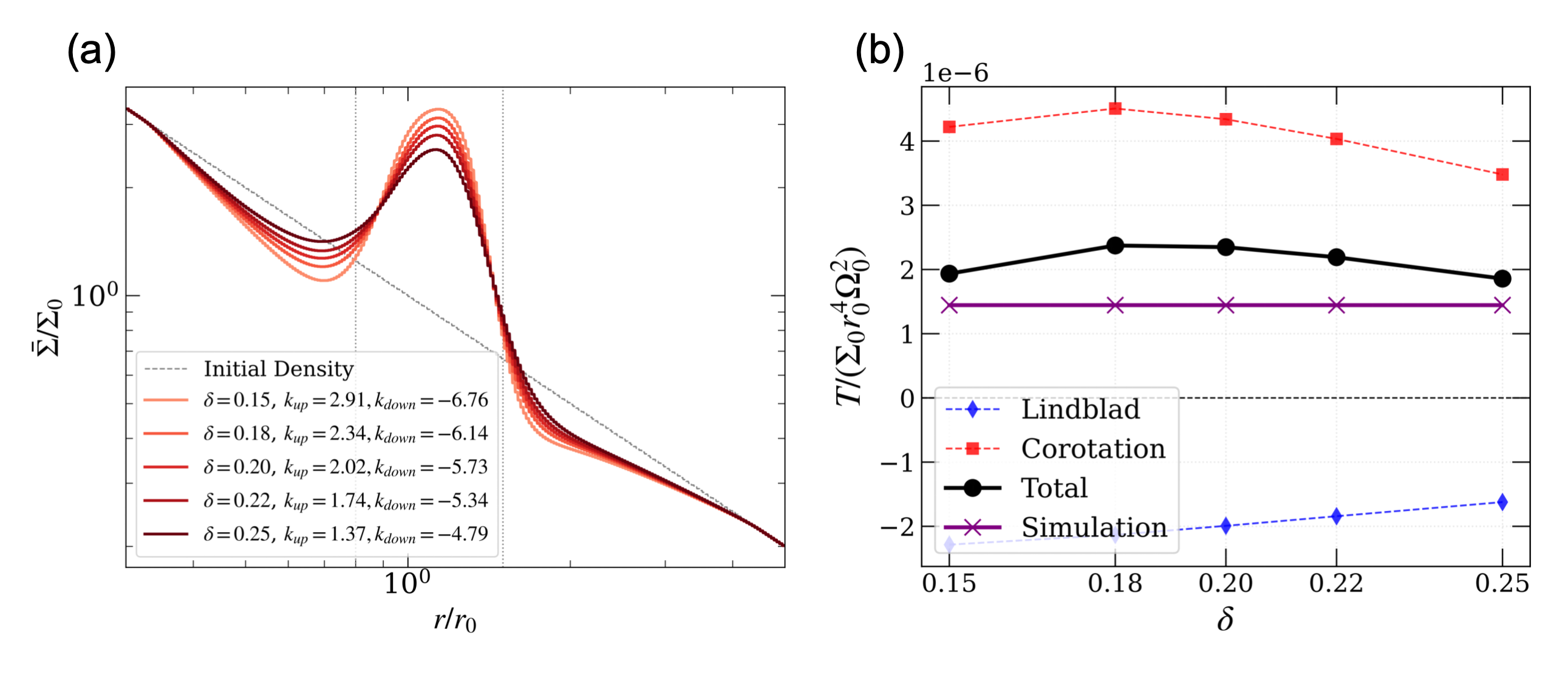}
    \caption{Theoretical calculations of the torque on a super-Earth for different disk surface density profiles. Here, we set $M_{\rm p} = 10M_{\oplus}$. The theoretical torque values are computed at $\Delta r_{\rm mid}/r_0 = -0.15$ in Fig.~\ref{fig:theory}. This model feature a dead zone with an inner boundary at $0.8 r_{0}$ and an outer boundary at $1.5 r_{0}$. Panel a illustrates the variation in the slope of the surface density profile (the power-law index) for different values of $\delta$ in Eq.~\ref{eq:alpha_profile}. Panel b displays the corresponding theoretical and simulated torques. The blue diamond-dashed line represents the Lindblad torque, and the red square-dashed line denotes the corotation torque. The black dot-solid line shows the sum of the two from analytical calculation, representing the total torque. The purple x-solid line corresponds to the total torque computed directly from Eq.~\ref{eq:Ttot_sim}.}
    \label{fig:slop_coro}
\end{figure*}

We acknowledge that corotation torques could be sensitive to the gas density gradient. Therefore, we vary the slope of the density transition in the model at $\Delta r_{\rm mid}/r_0 = -0.15$ in Fig.~\ref{fig:theory}. This model features a dead zone with an inner boundary at $0.8 r_{0}$ and an outer boundary at $1.5 r_{0}$. As shown in Fig. \ref{fig:slop_coro}(a), adopting different values of $\delta$ in Eq.~\ref{eq:alpha_profile} leads to variations in the slopes of the surface density profile (i.e., the power-law index). The slope of the ascending part ranges from 1.37 to 2.91, while the slope of the descending part ranges from $-4.79$ to $-6.76$. Fig. \ref{fig:slop_coro}(b) presents the theoretical calculations of the corotation torque and Lindblad torque, along with a comparison between the total theoretical torque and the simulation results. Although slight variations in the corotation torque are observed, it remains the dominant contribution to the total torque even when the surface density transition becomes shallower. 

\begin{figure*}
    \centering
    \includegraphics[width=0.95\linewidth]{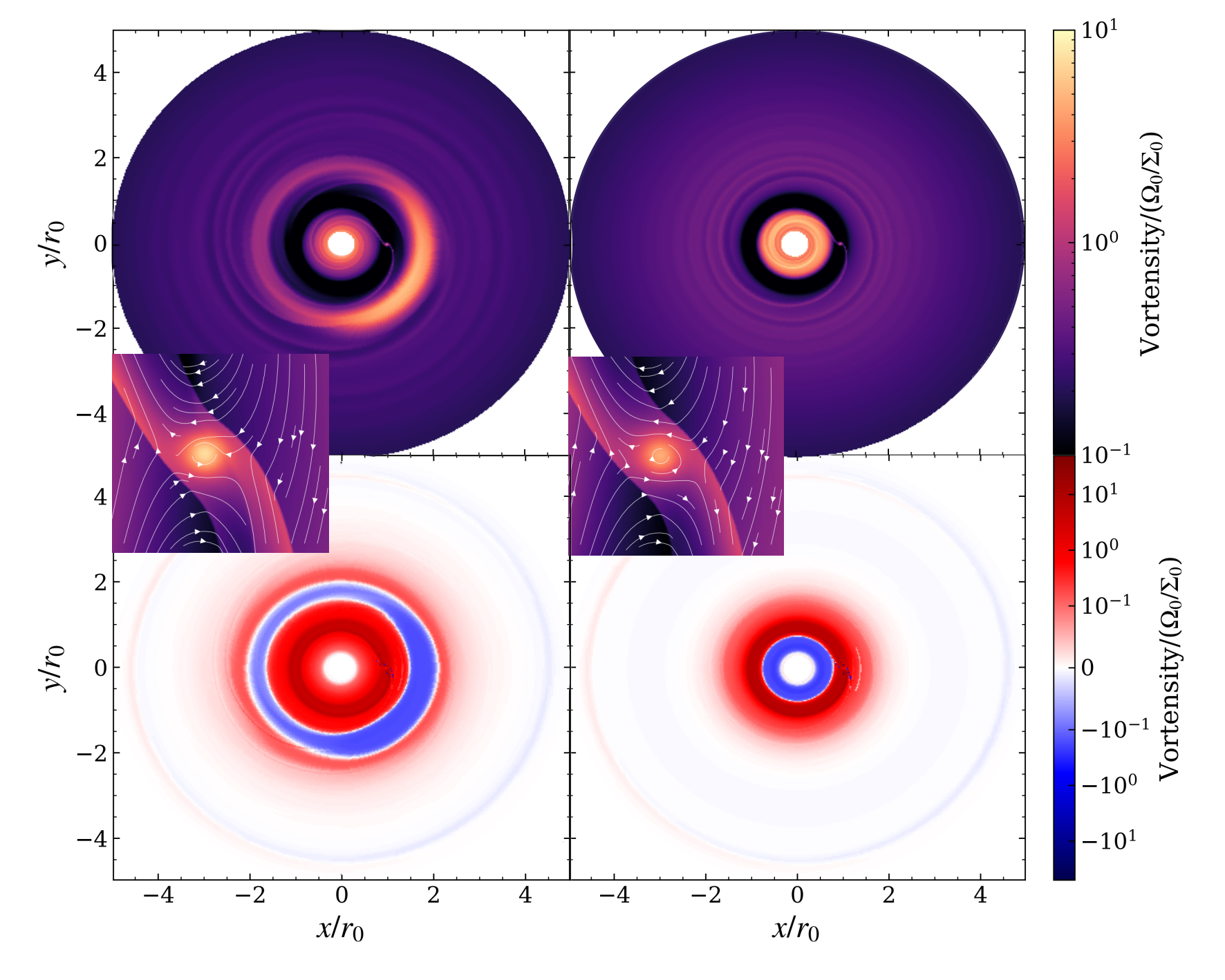}
    \caption{Pseudocolor maps of surface density and
    vortensity distribution. The left panels corresponds to the case with a vortex, while the right panels shows the case without a vortex. The inset in the middle shows a magnified view around the planet, with white streamlines depicting the velocity field. These correspond to the two locations marked by pentagram symbols in the lower-left and upper-right corners of Fig.\ref{fig:torque_scatter}(c), respectively. In the left panel, an asymmetric ring structure is visible outside the annular surface density bump, highlighting the presence of a vortex. 
    }
    \label{fig:vortex}
\end{figure*}

\subsection{Effect of Vortex}

In the super-Earth case, the setup is deliberately configured to suppress vortex formation. While in the Jupiter case, vortices form under certain conditions. As shown in Fig.~\ref{fig:torque_scatter}(c), the size of each scatter point represents the standard deviation of the torque time series, which reflects the intensity of vortices. Two representative points are selected: one in the lower-left corner (with vortex) and the other in the upper-right corner (without vortex) of Fig.~\ref{fig:torque_scatter}(c). Their corresponding vortensity distributions and velocity streamlines around the planet are displayed in Fig.~\ref{fig:vortex}. 

The formation of vortices in protoplanetary disks is primarily driven by instabilities that arise under specific disk conditions. 
Vortices can be generated through multiple mechanisms, including Rossby Wave Instability (RWI) at the edges of gaps carved by massive planets in low-viscosity disks, at the boundaries of accretionally inactive dead zones, via the influence of binary companions, or through baroclinic instability. 
The RWI \citep{lovelace1999, lihui2000, lihui2001,
Lyra2013} is a dominant mechanism responsible for generating the large-scale, dust-trapping vortices observed in transitional disks. The RWI is characterized as an ``edge mode'' instability, 
analogous to the Kelvin-Helmholtz instability, which operates at sharp pressure maxima within the disk. It efficiently converts excess shear into vorticity, leading to the formation of long-lived, coherent vortical structures \citep{balmforth2001, koller2003, lihui2005,Li2020}. 


In this study, vortex formation results from the combined effects of Jupiter's gap-opening process and the ring-shaped surface density bump. The gap carved by Jupiter produces an annular region of reduced surface density, while the annular surface density bump is an accumulation of mass. The relative position and interaction between these two features determine both the occurrence and strength of any resulting vortices. For vortices at dead zone edges is primarily driven by gas accumulates at this viscosity transition, to maintain a radially constant accretion rate (${\dot M}\sim \nu \Sigma_g$), leading to the formation of a local maximum in the gas pressure profile (a ``bump''). When this pressure bump is sufficiently sharp (with a width $\Delta_{\rm DZ}\lesssim 2H$, where $H$ is the local scale height), it becomes unstable to the RWI, which non-linearly evolves into long-lived, anticyclonic vortices \citep{Miranda2017}. This vortex generation mechanism at the dead zone edge is self-sustaining because the ongoing accretion flow continuously replenishes the gas at the bump, allowing the RWI and the vortices to persist for thousands of orbits.

As shown in the left panel of Fig.~\ref{fig:vortex}, pronounced vortices typically form near the outer edge of the dead zone (close to $r_{\mathrm{dz\_out}}$). This is attributed to the richer material reservoir and the longer boundary extent beyond the outer boundary compared to its inner region interior to the inner boundary, which promotes stronger material exchange across the dead zone interface. Such conditions readily establish the sharp surface density gradients required for the RWI to develop. The presence of Jupiter opens a gap in the protoplanetary disk. When the planet is located near the outer boundary of the dead zone, the gap it carves tends to smooth out the sharp surface density distribution induced by the dead zone, thereby disrupting the necessary conditions for the RWI and suppressing vortex formation. As a result, in Fig.~\ref{fig:torque_scatter}(c), configurations farther from the outer boundary of the dead zone (toward lower-left) are more prone to vortex generation. Conversely, cases near the outer boundary (toward upper-right) exhibit little to no vortices. Correspondingly, the vortex intensity in Fig.~\ref{fig:torque_scatter}(b) follows the trend: \ding{192}$\approx$\ding{193}$<$\ding{194}$<$\ding{195}.



\begin{figure*}
    \centering
    \includegraphics[width=0.9\linewidth]{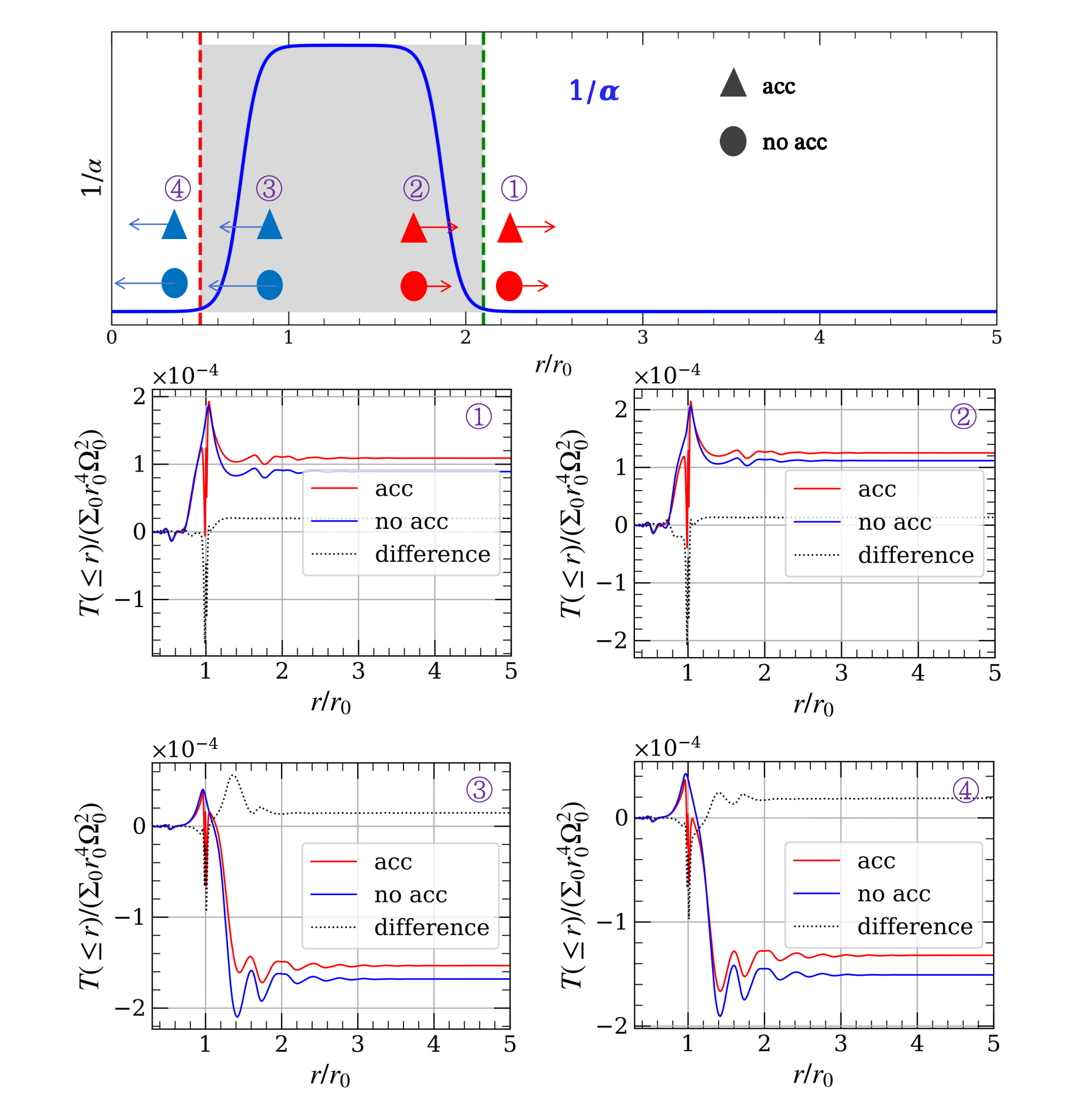}
    \caption{The effect of Jupiter's accretion on the total torque. The upper panel illustrates the overall trend, where triangular markers represent cases with accretion and circular markers represent cases without accretion. Blue indicates inward migration, and red indicates outward migration. The length of each arrow represents the relative magnitude of the torque. The four lower panels correspond to the regions marked by \ding{192} \ding{193} \ding{194} \ding{195} in the upper panel. Accretion provides a torque that promotes outward migration. 
    }
    \label{fig:acc}
\end{figure*}

\subsection{Effect of Planet Accretion}

For the migration of Jupiter, we conducted supplementary investigations that include planet accretion. A comparative analysis of accretion and non-accretion cases for the four regions near the dead zone boundary, labeled \ding{192}--\ding{195} in Fig.~\ref{fig:acc}, is presented in the upper panel of Fig.~\ref{fig:acc}. 

According to \cite{li2024}, when the disk's $\alpha$ parameter exceeds 0.003, accretion causes the planet to transition from inward to outward migration. In our setup, the dead zone has an $\alpha$ of $10^{-4}$, while the active region has an $\alpha$ of $10^{-2}$. The $\alpha$ in the active region exceeds the critical value required for accretion to exert an influence, whereas the $\alpha$ in the dead zone is insufficient for accretion to have a significant effect. Although accretion enhances the outward migration tendency of Jupiter, it is not sufficient to reverse the overall migration direction. 

As shown in the upper panel of Fig.~\ref{fig:acc}, the length and direction of the arrows represent the magnitude and direction of the torque governing Jupiter's migration.

In regions \ding{192}--\ding{195}, accretion exerts a discernible influence. Specifically, the outward migration trend is enhanced in regions \ding{192} and \ding{193}, while the inward migration trend is weakened in regions \ding{194} and \ding{195}.



\subsection{The Cases with a Surface Density Gap}

Besides the surface density bumps manifested as ringed structures, there are many annular gaps observed in high-resolution ALMA observations as mentioned above.  To determine whether the presence of a surface density gap in the disk is a result of giant planets forming at that specific location or, conversely, whether the gap itself directs planetary migration to that region, we investigate the migration directions of planets in a disk with a pre-existing surface density gap. 

\begin{figure*}
    \centering
    \includegraphics[width=0.9\linewidth]{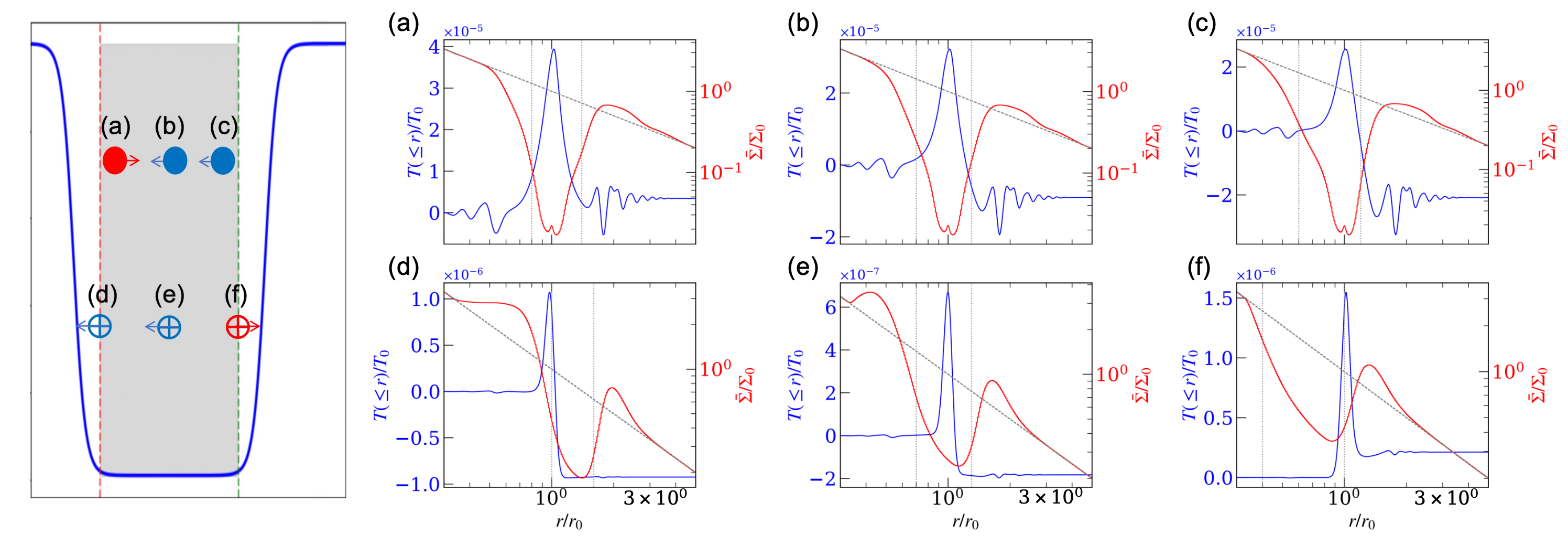}
    \caption{The migration behavior of planets in a disk with a surface density gap. We inverted the $\alpha$ profile by swapping the values in the originally defined dead and active zones, while maintaining a smooth transition between them. The top row of panels (panel a--c) displays the migration map for Jupiter, whereas the bottom row (panel d--f) corresponds to super-Earths. For the upper column (left to right): the panel a employs $r_{\mathrm{dz\_in}} = 0.8\,r_0$, $r_{\mathrm{dz\_out}} = 1.4\,r_0$; the panel b uses $r_{\mathrm{dz\_in}} = 0.7\,r_0$, $r_{\mathrm{dz\_out}} = 1.3\,r_0$; and the panel c is specified as $r_{\mathrm{dz\_in}} = 0.6\,r_0$, $r_{\mathrm{dz\_out}} = 1.2\,r_0$. For the lower column (left to right): the panel d adopts $r_{\mathrm{dz\_in}} = 1.0\,r_0$, $r_{\mathrm{dz\_out}} = 1.6\,r_0$; the panel e has $r_{\mathrm{dz\_in}} = 0.7\,r_0$, $r_{\mathrm{dz\_out}} = 1.3\,r_0$; and the panel f corresponds to $r_{\mathrm{dz\_in}} = 0.4\,r_0$, $r_{\mathrm{dz\_out}} = 1.0\,r_0$. 
    The selected configurations, indicating the relative positions between the surface density gap and the planet, are marked by diamond symbols in Fig.~\ref{fig:torque_scatter}(c) and (d). The resulting migration directions, as shown in the left panel, are opposite to those shown in Fig.~\ref{fig:torque_scatter}(b). 
    }
    \label{fig:reverse}
\end{figure*}

We employ an $\alpha$ profile that is inverted relative to Fig.~\ref{fig:torque_scatter}(a), i.e. an $\alpha$ bump, to generate a surface density gap in the disk (Fig.~\ref{fig:reverse}). Specifically, for Jupiter, the annular shaded region in Fig.~\ref{fig:torque_scatter}(a) is assigned as the high viscosity zone with $\alpha = 10^{-2}$, while the remaining regions are designated as the low viscosity zone with $\alpha = 10^{-4}$, with a smooth transition between them implemented via Eq.~\ref{eq:alpha_profile}. For super-Earths, a similar configuration is employed, except that for the high viscosity region $\alpha$ is set to $0.001$. 

As shown in Fig.~\ref{fig:reverse}, panels a--c present the migration behavior of Jupiter in a disk with an annular surface density gap, while panels d--f show the corresponding results for super-Earths. The relative positions of the planets with respect to the surface density gap in these configurations are marked by diamond symbols in Fig.~\ref{fig:torque_scatter}(c) and (d). For Jupiter, when located within the ring-shaped surface density gap, its migration direction is opposite to that observed in the annular surface density bump: with respect to the center of the surface density gap, planets migrate outward in the interior region and inward in the exterior region, causing Jupiter to converge toward the gap center. Moreover, gap opening by Jupiter further deepens and reinforces the pre-existing intrinsic surface density gap. For super-Earths, planets inside the ring-shaped surface density gap predominantly undergo inward migration. However, on or slightly outside the outer boundary of the surface density gap, outward migration occurs, whereas inward migration persists near the inner boundary. This implies that super-Earths are unlikely to remain trapped within the ring-shaped surface density gap, suggesting that intrinsic surface density gaps may not harbor super-Earths. 

Observational surveys often target surface density gaps in protoplanetary disks as potential sites for planet detection, including Jupiter-mass and super-Earth-mass planets. Our study, however, suggests that intrinsic surface density gaps may not be able to retain super-Earths, whereas Jupiters can not only reside in such gaps but also further deepen and reinforce them through gap-opening processes.

\subsection{Implementation of Artificial Azimuthal Torque}\label{sec:torque_bump}

Recent advances in migration theory demonstrate that migration behavior is dependent on the $\alpha$ viscosity parameter \citep{Paardekooper2023}. The use of an $\alpha$ dead zone to mimic surface density bumps and gaps carries the risk of neglecting the potential effects of vortices on planetary migration. Therefore, we also employed an alternative method as a supplementary approach, applying an artificial torque to the protoplanetary disk to generate a surface density bump instead of an $\alpha$ dead zone.

To investigate the disk's response to localized angular momentum redistribution, we implement an artificial torque function $T_{\rm art}(r)$ centered at a fiducial radius $r_{\rm mid}$. The torque profile is designed to converge material toward $r_{\rm mid}$, creating a density bump (or gap, depending on the sign) at this location. 

The artificial torque is defined as:
\begin{equation}
T_{\rm art}(r) = \begin{cases}
-T_0 \sin(x) \times 10^{-5}, & \text{if } |r - r_{\rm mid}| \leq W \\
0, & \text{otherwise}
\end{cases}
\end{equation}
where $T_0$ is the torque amplitude, $W$ is the effective width of the torque application region, and the normalized radial coordinate $x$ is given by:
\begin{equation}
x = \frac{r - r_{\rm mid}}{W} \pi
\end{equation}

The sinusoidal form ensures that the torque smoothly vanishes at the boundaries ($|r - r_{\rm mid}| = W$) while making the surface density profile reaching maximum strength at $r = r_{\rm mid}$.  
The negative sign in the expression ensures a converging torque profile: for $r < r_{\rm mid}$ (where $x < 0$), $-\sin(x) > 0$, producing a positive (outward-directed) torque that pushes material outward toward $r_{\rm mid}$; conversely, for $r > r_{\rm mid}$ (where $x > 0$), $-\sin(x) < 0$, producing a negative (inward-directed) torque that pulls material inward toward $r_{\rm mid}$.


\begin{figure*}
    \centering
    \includegraphics[width=0.9\linewidth]{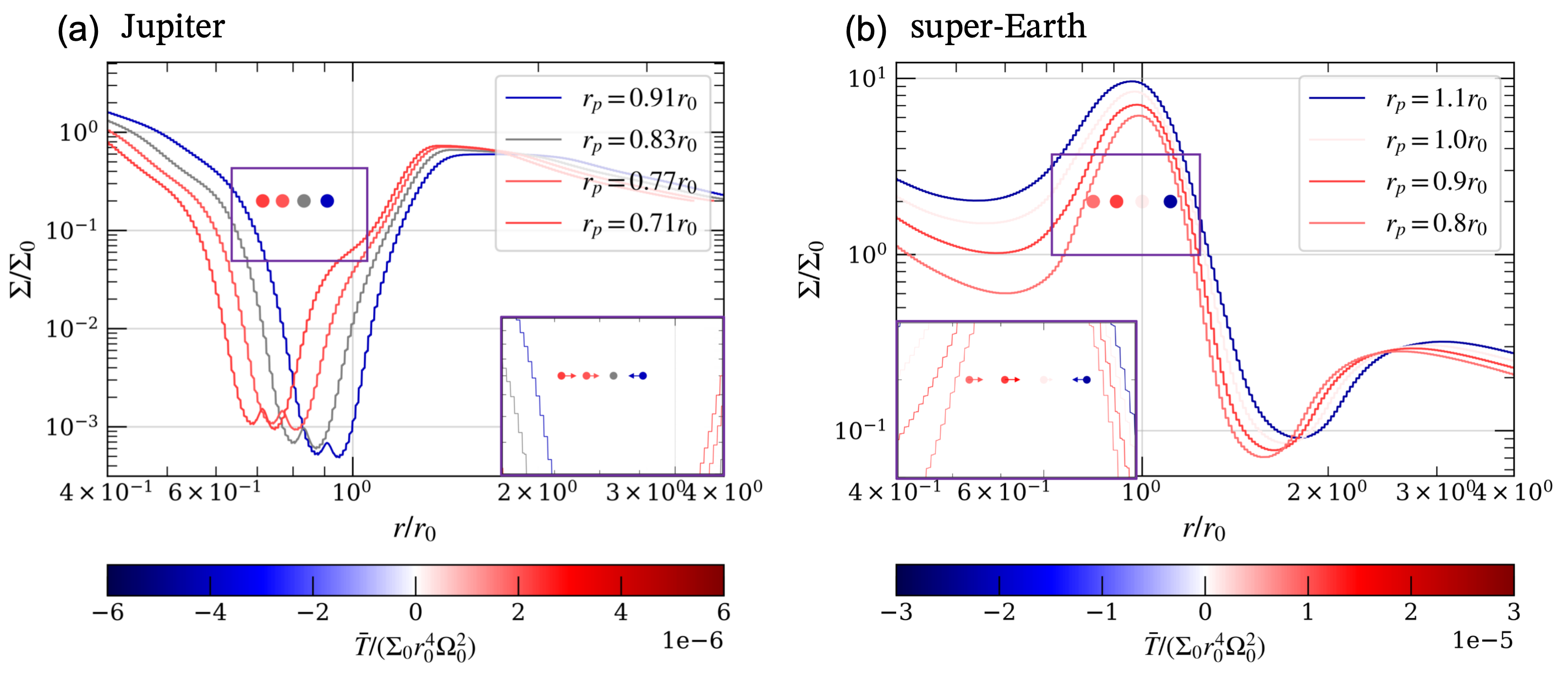}
    \caption{The surface density profile derived from the artificial torque method. The solid line represents the surface density, and circular markers indicate the planet's location. The color bar shows the magnitude of the average torque exerted on the planet for the given surface density profile and planetary position. The insets in the lower-left and lower-right corner, which zoom in on the region enclosed by the purple box, illustrate the direction of planetary migration. Panel a illustrates the Jupiter case, in which the planet migrates into the surface-density gap. Panel b shows the super-Earth case, where the planet moves toward the surface-density bump. }
    \label{fig:artificial}
\end{figure*}

We examine four representative cases for Jupiter migration around a surface density gap. The Jupiter is placed at $r = 1\ r_{0}$, with a gap width $W = 0.5\ r_{0}$, amplitude $T_0 = 5 \Sigma_0 r^4_0 \Omega^2_0$, and mid-line locations $r_{\mathrm{mid}}$ set to $1.4 r_{0}$, $1.3 r_{0}$, $1.2 r_{0}$, and $1.1 r_{0}$, respectively. Each configuration is then rescaled so that its mid-line coincides with $r = r_0$. Consequently, the planet positions in these four cases become $r_{\mathrm{p}} = 0.91 r_0$, $r_{\mathrm{p}} = 0.83 r_0$, $r_{\mathrm{p}} = 0.77 r_0$, and $r_{\mathrm{p}} = 0.71 r_0$, as illustrated in Fig.~\ref{fig:artificial}(a). The solid line traces the surface-density profile, and a circular marker indicates the planet location. The color bar encodes the magnitude of the time-averaged torque experienced by the planet for the given density structure and orbital radius. An inset in the lower-left corner, zooming into the region outlined by the purple box, illustrates the direction of planetary migration. 
From the computed average torques in these four cases, we confirm that Jupiter indeed converge toward the surface density gap, as shown by the migration directions in the inset of Fig.~\ref{fig:artificial}(a).

We also examine four representative cases for super-Earth migration around a surface density bump. The super-Earth is placed at $r = 1\ r_{0}$, with a bump width $W = 1\ r_{0}$, amplitude $T_0 = 10 \Sigma_0 r^4_0 \Omega^2_0$, 
and mid-line locations $r_{\mathrm{mid}}$ set to $0.9 r_{0}$, $1.0 r_{0}$, $1.1 r_{0}$, and $1.2 r_{0}$, respectively. Each configuration is then rescaled so that its mid-line coincides with $r = r_0$. Consequently, the planet positions in these four cases become $r_{\mathrm{p}} = 1.1 r_0$, $r_{\mathrm{p}} = 1.0 r_0$, $r_{\mathrm{p}} = 0.9 r_0$, and $r_{\mathrm{p}} = 0.8 r_0$, as illustrated in Fig.~\ref{fig:artificial}(b). We also concluded that super-Earth converge toward the surface density gap.

\section{Conclusion and Discussions}
\label{sec:discussion}


In this study, we examine the migration behavior of planets with varying masses embedded at different initial locations in a protoplanetary disk containing a rim-like structure using hydrodynamical simulations with \texttt{Athena++}. This rim-like structure is generated by a viscosity transition profile.
Our primary goal is to identify the preferred detection regions for Jupiter and super-Earth mass planets. We show that hot Jupiters undergo outward migration from the center of the gas bump, dominated by Lindblad torques, with additional contributions from accretion processes. Meanwhile, super-Earths exhibit convergent migration toward the interior of the gas bump from both sides, often becoming trapped near its center, as supported by both numerical simulations and theoretical analyses. This result suggests that the super-Earth could be preferred to be detected around the surface density bump rather than the surface density gap in the structured PPDs.
Furthermore, for a disk with a surface density gap, we find that the Jupiter mass planet tends to be trapped in the gap center, which in turn modifies the preexisting gap profile by the tidal torque. Thus caution should be exercised when using the gap profile to constrain the properties of the unseen planets.

Our calculations are scale-free; however, we can provide an example application to observational estimates. For instance, the cumulative torque in the top-left panel of Fig.~\ref{fig:density_r_torque_diff} is approximately $T\sim 10^{-4} \Sigma_0 r_0^4 \Omega_0^2$. 
We adopted typical ALMA-observed locations and densities of rings to estimate the migration timescale \citep{Andrews2018, Yoshida2025}. By adopting an orbital radius of $r_0 = 50 \, \text{AU} = 50 \times 1.496 \times 10^{13} \, \text{cm}$, a surface density of $\Sigma_0 = 20 \, \text{g/cm}^2$, we can estimate the migration rate for a Jupiter mass planet as $\dot{r}_{\rm p}/r_{\rm p}={2T}/{M_{\rm p} \Omega_{\rm p} r_{\rm p}^{2}}\simeq 1.1 \times 10^{-3}\Omega_{\rm p}$, which corresponds to a migration timescale of $\simeq 5 \times 10^4\ {\rm yrs}$. The migration timescale for the super-Earth is also on the same order of magnitude. Such a short timescale suggests an efficient process for the trapping of the planets within the rim. 




It should be noted that this study has several limitations. For instance, in our hydrodynamical simulations, the planetary orbits were held fixed to compute the total torque and infer the direction of migration, rather than allowing the planets to migrate freely. Consequently, the actual evolution of the torque during migration remains unexplored. Furthermore, due to computational constraints, our investigation was limited to single-planet systems. The effects of multi-planet interactions, such as gravitational scattering and resonant capture, were not considered and warrant further study.

\acknowledgements
We thank the anonymous referee for their valuable suggestions and comments, which have significantly improved the manuscript. This work is partly supported by the National Natural Science Foundation of China (Grant No. 12133005, Z.Y.C. and S.M.). Y.P.L is supported in part by the National Natural Science Foundation of China (Grants NO. 12373070, and 12192223), the National Natural Science Foundation of Shanghai (Grant NO. 23ZR1473700). The authors acknowledge the High-Performance Computing platform at Westlake University for providing computational and data storage resources that have contributed to the results in this paper. The calculations have made use of the High Performance Computing Resource in the Core Facility for Advanced Research Computing at Shanghai Astronomical Observatory.

\bibliography{citation.bib}{}
\bibliographystyle{aasjournalnolink}

\end{CJK*}
\end{document}